\documentclass[useAMS,usenatbib]{mn2e}
\usepackage{amsmath}
\usepackage{amscd}
\usepackage{amssymb}
\usepackage{txfonts}
\RequirePackage{color}

\allowdisplaybreaks[1]

\title[Planet X revamped after the discovery of the Sedna-like object 2012 VP$_{113}$?]{Planet X revamped after the discovery of the Sedna-like object 2012 VP$_{113}$?}

\author[L. Iorio]{L.
Iorio$^{1}$\thanks{E-mail:
lorenzo.iorio@libero.it}\\
$^{1}$I Ministero dell'Istruzione, dell'Universit\`{a} e della
Ricerca (M.I.U.R.), Viale Unit\`{a} di Italia 68
Bari, (BA) 70125,
Italy}

\begin{document}

\maketitle

\label{firstpage}

\begin{abstract}
The recent discovery of the Sedna-like dwarf planet 2012 VP$_{\rm 113}$ by Trujillo and Sheppard has revamped the old-fashioned hypothesis that a still unseen trans-Plutonian object of planetary size, variously dubbed over the years as Planet X, Tyche, Thelisto,  may lurk in the distant peripheries of the Solar System. This time, the presence of a super-Earth with mass $m_{\rm X} = 2-15m_{\oplus}$ at a distance $d_{\rm X}\approx 200-300$ astronomical units (AU) was proposed to explain the observed clustering of the arguments of perihelion $\omega$ near $\omega \approx 0^{\circ}$ but not $\omega\approx 180^{\circ}$  for Sedna, 2012 VP$_{\rm 113}$ and other minor bodies of the Solar System with perihelion distances $q>30$ AU and semimajor axes $a>150$ AU. Actually, such a scenario is strongly disfavored by the latest constraints $\Delta\dot\varpi$ on the anomalous perihelion precessions of some Solar System's planets obtained with the INPOP and EPM ephemerides. Indeed, they yield $d_{\rm X}\gtrsim 496-570$ AU ($m_{\rm X}=2m_{\oplus}$), and $d_{\rm X}\gtrsim 970-1111$ AU ($m_{\rm X} = 15 m_{\oplus}$). Much tighter constraints could be obtained in the near future from the New Horizons mission to Pluto.
\end{abstract}

%\centerline
%{Keywords: Oort Cloud--Kuiper belt: general--celestial mechanics--ephemerides}

%

\begin{keywords}
Oort Cloud--Kuiper belt: general--gravitation--celestial mechanics--ephemerides
\end{keywords}
\section{Introduction}
Since the early work of \citet{Lowell1915}, the hypothesis that the remote peripheries of the Solar System may host a still unseen major body is regularly re-emerged over the years (see, e.g., \citep{1984PNAS...81..801R, 1984Natur.308..713W, 2006Icar..184..589G, 2008AJ....135.1161L, 2009MNRAS.400..346I, 2010MNRAS.407L..99M, 2011Icar..211..926M, 2011ApJ...726...33F, 2012MEEP....1..121L, 2012DDA....43.0501G, 2012CeMDA.112..117I, 2013CeMDA.116..357I, Nat2014}), supported by a number of more or less sound indirect observational motivations  ranging from alleged periodicities detected in paleontological fossil records on the Earth \citep{2013ApJ...773....6M} to certain morphological features of the Edgeworth-Kuiper belt \citep{2012MEEP....1..121L}. Such a hypothetical trans-Plutonian object has been variously dubbed so far as Planet X \citep{Lowell1915}, Nemesis \citep{1984PNAS...81..801R}, Tyche \citep{2011Icar..211..926M}, Thelisto \citep{2013CeMDA.116..357I}; in the following, we will refer to it as PX.

A direct imaging search for a distant companion to the Sun with the Wide-field Infrared Survey Explorer (WISE) spacecraft-based mission has recently yielded negative results, at least as far as gaseous giants are concerned \citep{2014ApJ...781....4L}. Indeed, bodies with the physical characteristics of Saturn and Jupiter cannot exist at less than 28000 and 82000 astronomical units (AU), respectively \citep{2014ApJ...781....4L}, while the closer distance limit for a Jupiter-mass brown dwarf is 26000 AU \citep{2014ApJ...781....4L}.

Nonetheless, the latest hint about the possible existence of a remote planetary-sized object in the outskirts of the Solar System came recently from the discovery of the Sedna-like dwarf planet 2012 VP$_{113}$ \citep{Nat2014}. Indeed,
\citet{Nat2014} remarked that the clustering of the arguments of pericenter $\omega$ near 0$^{\circ}$ but not 180$^{\circ}$ for Sedna, 2012 VP$_{113}$  and other extreme Solar System's bodies with perihelion distances $q>30$ AU and semimajor axes $a> 150$ AU could not be attributed to an observational bias effect. As a viable candidate to explain such a pattern, \citet{Nat2014} suggested the possible existence of a still unseen distant perturber  in the form of a super-Earth ($m_{\rm X}=2-15 m_{\oplus}$) rock-ice planet moving in a circular, low inclination orbit between 200 and 300 AU. More precisely, numerical simulations showed that, while the presently known mass distribution of the Solar System would produce an inner Oort cloud with randomly distributed perihelia, a pointlike object as the one just mentioned would be able to let the perihelia of the inner Oort cloud objects librate around $\omega=0^{\circ}\pm 60^{\circ}$ over time scales as long as billions of years \citep{Nat2014}.
%Alternatively, a Neptune-sized object in a highly inclined trajectory at about 1500 AU was considered as well.
Different orbital configurations for such a putative disturbing body were considered in \citep{Nat2014}. Concerning the direct observability of the hypothesized perturber of terrestrial type, \citet{Nat2014}, who did not mention the all-sky WISE survey by \citet{2014ApJ...781....4L},  remarked that if the albedo of a super-Earth  at 250 AU were low enough, it would escape from detection in current all-sky surveys \citep{2011AJ....142...98S}.

For other motivations supporting a PX scenario, see the Introduction of \citet{2013CeMDA.116..357I} and references therein.
Among them, we mention that the dynamical action of a distant, isolated point-like mass can mimic the impact on Solar System's major bodies of a specific kind of the subtle External Field Effect (EFE) arising in the framework of the MOdified Newtonian Dynamics (MOND) \citep{2009MNRAS.399..474M, 2010AIPC.1241..935I, 2011MNRAS.412.2530B, 2014arXiv1402.6950H}.
\section{Updated constraints from the planetary perihelion precessions}
%
%The appealing hypothesis by \citet{Nat2014} must be put on the test in view of the latest constraints on the location of a hypothetical %trans-Plutonian object of planetary size inferred both directly from imaging surveys \citep{2014ApJ...781....4L} and indirectly from the dynamics of Solar %System's major bodies \citep{2012CeMDA.112..117I}.
%
In his recent analysis of the all-sky survey with the WISE data, \citet{2014ApJ...781....4L} inferred tight lower bounds on the distance $d_{\rm X}$ of putative far giant Saturn-like and Jupiter-like planets: $d_{\rm X} \gtrsim 28000$ AU for $m_{\rm X}=m_{\rm Sat}$, and $d_{\rm X} \gtrsim 82000$ AU for $m_{\rm X}=m_{\rm Jup}$, respectively. Moreover, he found that a Jupiter-mass brown dwarf cannot be located at less than 26000 AU. Actually, his analysis did not deal with rock-ice terrestrial planets as the one postulated by \citet{Nat2014}.

Model-independent, dynamical constraints on $d_{\rm X}$ were inferred by \citet{2012CeMDA.112..117I} for different values of $m_{\rm X}$ from the upper bounds $\Delta\dot\varpi$ on the anomalous secular precessions of the longitude of perihelion $\varpi$ of some known planets of the Solar System computed with earlier versions of the INPOP ephemerides \citep{2010IAUS..261..159F}. \citet{2012CeMDA.112..117I} preliminarily obtained $d_{\rm X}\gtrsim 250-450$ AU for $m_{\rm X} = 0.7 m_{\oplus}$. Since it is $d_{\rm X}\propto \left(m_{\rm X}/\Delta\dot\varpi\right)^{1/3}$ \citep{2012CeMDA.112..117I}, it is straightforward to refine such estimates by using the latest results on $\Delta\dot\varpi$ obtained with more recent planetary ephemerides \citep{2011CeMDA.111..363F, 2013MNRAS.432.3431P}. From Figure 1 of  \citet{2012CeMDA.112..117I}, it can be noticed that, for a given value of $m_{\rm X}$ and by keeping the ecliptic latitude $\beta_{\rm X}$ fixed to some low values, the combined use of the perihelia of Earth, Mars and Saturn allows to constrain effectively $d_{\rm X}$ for practically all values of the ecliptic longitude $\lambda_{\rm X}$. In the case $m_{\rm X} = 0.7 m_{\oplus}$  \citep{2012CeMDA.112..117I}, the values by \citet{2013MNRAS.432.3431P} for the perihelion precessions of Mars and Saturn provide us with overall tighter bounds of the order of $d_{\rm X}\gtrsim 350-400$ AU. In turn, such revised bounds can be easily extended to the scenario proposed by \citet{Nat2014}; it turns out that $d_{\rm X}\gtrsim 496-570$ AU for $m_{\rm X}=2m_{\oplus}$ and $d_{\rm X}\gtrsim 970-1111$ AU for $m_{\rm X} = 15 m_{\oplus}$,  respectively. Thus, the super-Earth scenario suggested in  \citep{Nat2014} to explain the perihelion clustering of the known objects with $q> 30$ AU and $a>150$ AU faces serious observational challenges.

Finally, it is worthwhile noticing that the New Horizons spacecraft en route to the Pluto system should be able to put even tighter limits on the location of a putative trans-Plutonian object \citep{2013CeMDA.116..357I}. Indeed, by assuming spacecraft's range residuals as little as 10 m, it should be possible to constrain the location of a rock-ice planet with $m_{\rm X} = 0.7 m_{\oplus}$ down to about $d_{\rm X} \gtrsim 4700$ AU \citep{2013CeMDA.116..357I}.
\section{Summary and conclusions}
The hypothesis of a trans-Plutonian super-Earth ($m_{\rm X}=2-15m_{\oplus}$) near the ecliptic at distances $d_{\rm X}\approx 200-300$ AU put forth after the discovery of 2012 VP$_{113}$ by Trujillo and Sheppard to explain the observed pattern of the perihelia of the Solar System's objects with perihelion distances $q>30$ AU and semimajor axes $a>150$ AU is ruled out by the current bounds $\Delta\dot\varpi$ on the anomalous secular perihelion precessions of some known planets of the Solar System.

Indeed, latest determinations of $\Delta\dot\varpi$ by Pitjeva and Pitjev with the EPM ephemerides yield $d_{\rm X}\gtrsim 496-570$ AU for $m_{\rm X}=2m_{\oplus}$, and $d_{\rm X}\gtrsim 970-1111$ AU for $m_{\rm X} = 15 m_{\oplus}$.

The New Horizons spacecraft, currently en route to Pluto, should allow to constrain the distance of a putative body with $m_{\rm X} = 0.7m_{\oplus}$ down to $d_{\rm X}\gtrsim 4700$ AU.

\bibliography{innerOortbib}{}
%-----------------------------------------

\end{document}